%% file: main.tex
\documentclass[%
superscriptaddress,
reprint,
preprintnumbers,
bibnotes,
 amsmath,
 amssymb,
 aps,
pra,
]{revtex4-1}
\usepackage{import}

\usepackage{natbib}
\usepackage{graphicx}
\usepackage{bmpsize}
\usepackage{dcolumn}
\usepackage{bm}
\usepackage{hyperref}
\hypersetup{colorlinks=true}
\usepackage{multirow}
\usepackage[dvipsnames]{xcolor}
\pdfoutput=1

\newcommand{\phystakingeng}{$-0.24$}

\begin{document}

\preprint{APS/123-QED}

\title{Modeling student pathways in a physics bachelor's degree program}

\author{John M. \surname{Aiken}}
\affiliation {Center for Computing in Science Education \& Department of Physics, University of Oslo, N-0316 Oslo, Norway}
\affiliation {Department of Physics and Astronomy, Michigan State University, East Lansing, Michigan 48824}

\author{Rachel Henderson}
\affiliation {Department of Physics and Astronomy, Michigan State University, East Lansing, Michigan 48824}
\author{Marcos D. \surname{Caballero}}
\affiliation {Center for Computing in Science Education \& Department of Physics, University of Oslo, N-0316 Oslo, Norway}
\affiliation {Department of Physics and Astronomy, Michigan State University, East Lansing, Michigan 48824}
\affiliation {CREATE for STEM Institute, Michigan State University, East Lansing, Michigan 48824}

\date{\today}

\begin{abstract}

Physics education research has used quantitative modeling techniques to explore learning, affect, and other aspects of physics education. However, these studies have rarely examined the predictive output of the models, instead focusing on the inferences or causal relationships observed in various data sets. This research introduces a modern predictive modeling approach to the PER community using transcript data for students declaring physics majors at Michigan State University (MSU). Using a machine learning model, this analysis demonstrates that students who switch from a physics degree program to an engineering degree program do not take the third semester course in thermodynamics and modern physics, and may take engineering courses while registered as a physics major. Performance in introductory physics and calculus courses, measured by grade as well as a students' declared gender and ethnicity play a much smaller role relative to the other features included the model. These results are used to compare traditional statistical analysis to a more modern modeling approach.

\end{abstract}

\maketitle

\section{\label{sec:intro}Introduction}

\import{sections/}{intro.tex}

\section{\label{sec:background}Background}
\import{sections/}{background.tex}

\section{\label{sec:data}Data \& Context}
\import{sections/}{figure1.tex}
\import{sections/}{data_based_on_figure.tex}

\section{\label{sec:methods}Methods}
\import{sections/}{featuretable.tex}

\import{sections/}{methods.tex}

\section{\label{sec:results}Results}
\import{sections/}{figure2.tex}

\import{sections/}{figure3.tex}

\import{sections/}{figure4.tex}

\import{sections/}{figure5.tex}

\import{sections/}{results.tex}

\section{\label{sec:discussion}Discussion of Results}
\import{sections/}{discussion.tex}

\section{\label{sec:limitations}Limitations and implications}
\import{sections/}{limitations.tex}

\section{\label{sec:methodsdiscussion}discussion of methods}

\import{sections/}{methods_discussion.tex}

\section{\label{sec:conclusion}Conclusion}
\import{sections/}{conclusions.tex}
\section{Acknowledgements}

This project was supported by the Michigan State University College of Natural Sciences including the STEM Gateway Fellowship, the Association of American Universities, the Olav Thon Foundation, and the Norwegian Agency for Quality Assurance in Education (NOKUT), which supports the Center for Computing in Science Education. Additionally, we would like to thank Gerald Feldman who suggested the concept of the sliding window analysis.

\bibliographystyle{apsrev4-1}
\bibliography{pathways}{}

\end{document}

%% file: intro.tex
Physics has long built data driven models to explain and to understand systems of study and Physics Education Research (PER) is without exception. In PER, these models have been used to explore learning outcomes \cite{ding2014uncovering}, understand career choice motivations \cite{dabney2014comparative}, and explore the use of different instructional strategies \cite{henderson2012use}, amongst many other topics. Generally within the PER literature, models are typically within the family of linear models (e.g., ordinary least squares, logistic regression) and are often evaluated by their ability to explain results through odds ratios and $p$-values as well as by using goodness-of-fit tests \cite{ding2012getting}. These type of methods are adequate when the goal is to explain and/or describe the data collected; however, it becomes difficult to reproduce such results when extending beyond the local setting \cite{shmueli2010explain}. Assessing the predictive output of these models can be one way that PER can begin to produce data driven models that can be compared and tested across different settings \cite{shmueli2010explain}.

As the field of PER continues to develop and refine its approaches to quantitative modeling, it is important to discuss the nature of modeling as well as the limitations and affordances of different approaches. Much of the work in PER assesses model fit without assessing the model's predictive power (e.g., \cite{ding2014uncovering, dabney2014comparative, henderson2012use}). These more traditional approaches tend to evaluate the fit of those models in the context of the collected data. In the current work, we employ a machine learning approach that emphasizes the generalizable nature of a quantitative model by first fitting the model to a data set and then separately evaluating the quality of the model using sequestered data, or "hold out" data, for which the model was not fit. This approach of assessing the predictive output of a model provides a direct quantifiable ``performance measure'' that can be used to compare multiple models or models from different data sets \cite{bradley1997use, hanley1982meaning}. This approach can allow model predictions to be compared across different settings where the models are attempting to predict the same outcome.

Rather than discussing the employed machine learning technique and abstractly comparing it to more common modeling approaches, here we will demonstrate the affordances and limitations of our approach through a specific case of investigating students who stay in the physics program compared to students who leave physics for engineering.
With the aim of introducing a new analysis method into the PER literature, we will analyze the predictions from models that are fit to university registrar data. Namely, we will examine the factors that impact a student's choice to switch from a physics to engineering degree program (intra-STEM switchers). In this study, the registrar data from Michigan State University (MSU) was previously investigated in \citet{aiken2016methods}; this type of data is typically collected by many institutions. Through this paper, we will introduce an algorithm for classification (random forest \cite{breiman2001random}) and compare this modeling approach to a summary statistic approach of using contingency tables and effect sizes. While the overarching focus of this paper is to compare these two approaches in a given research context, we do find that our work challenges the previous results stated in \citet{aiken2016methods}. We will demonstrate that the single most important component for remaining in the physics major is taking the third semester modern physics course.

For faculty who have spent any time advising physics majors, it might seem like this result is intuitive and it is true that this result served as an assumption in the previous work that analyzed student pathways in physics \cite{rodriguez2016gender}. It is precisely this intuition that allows us to focus on a new approach of modeling university registrar data and discuss the affordances and limitations of such an approach. Furthermore, although this result may be intuitive, it has gone unreported and thus, this work can provide the basis for future pathway studies.

As we use registrar data to illustrate what we might learn from this modeling approach, it is worth noting that this approach opens additional research questions that are worth reporting. Much of the literature that examines the pathways of STEM students focus on students who leave STEM completely \cite{seymour2000talking, chen2013stem} and not on those who leave one STEM discipline for another. Because at least one third of the students registering as physics majors at MSU earn degrees in engineering \cite{aiken2016methods}, through this work, we can investigate the following three questions related to whether students stay in a physics bachelor's program or leave for an engineering degree:

\begin{enumerate}
  \item Factors that describe students who leave STEM are well-documented. Some of those factors are present in university registrar data either directly or through various proxies. Which of the factors identified in the literature impact students to remain a physics major or leave the physics bachelor's program for an engineering degree?
  \item In prior work, \citet{aiken2016methods} found that performance metrics between physics and engineering graduates differed. What were the effect sizes of these performance factors and how do they impact models that compare the effect of various factors against each other?
  \item As our goal is to understand intra-STEM switchers, what did we learn about students who register for physics but leave for engineering?
\end{enumerate}

This paper is organized as follows. In Section \ref{sec:background}, we present prior work on STEM pathways to document specific features that we might be able to find in the MSU registrar data (Section \ref{sec:data}). We discuss the traditional and machine learning approaches in Section \ref{sec:methods} and refer the reader to \citet{young2018using} for additional details on the random forest technique. We then present the models tested (Section \ref{sec:allthemodels}) as well as the resulting output and validation (Section \ref{sec:results}). We then discuss the findings, in the context of our data to demonstrate what might be gained from using this modern approach (Section \ref{sec:discussion}) as well as the specific limitations of our work (Section \ref{sec:limitations}). Finally, we critique the different modeling approaches, discussing what affordances and limitations we find more generally (Section \ref{sec:methodsdiscussion}) and offer some concluding remarks (Section \ref{sec:conclusion}).

%% file: background.tex
The previous two decades have seen a large increase in student enrollment in STEM including physics bachelor's degree programs \cite{national2013adapting}. These large changes have been driven by a variety of efforts both nationally and locally including more aggressive recruiting of students in STEM majors and better retention of current majors in STEM programs. Much of this work was informed by research on why students leave STEM majors \cite{seymour2000talking, chen2013stem, tai2001gender}. While there is substantial and continued research into how and why students leave STEM, there is little understanding surrounding why students might leave a particular STEM field (such as physics) for another STEM field. Understanding these intra-STEM switchers can help physics departments explain attrition rates as well as assist departments in developing a better of understanding of how they are (or are not) meeting the needs of their current and potential majors. Furthermore, physics departments can directly benefit from understanding why students leave or stay in their programs. These data can be used to advocate for curricular changes, new learning environments, and up-to-date teaching practices if their undergraduate program is not meeting the department's desires.

\subsection{Leaving STEM Literature}
Leaving or switching from STEM to other majors has been explored in many contexts: educational, sociological, and through a Discipline Based Education Research (DBER) lens. Results from these studies have continually identified three reasons why students leave STEM: lack of interest \cite{seymour2000talking, chen2013stem, marra2012leaving}; poor performance \cite{griffith2010persistence,chen2013stem,shaw2010patterns,aiken2016methods, hackett1989exploration}; and differential experiences among groups \cite{seymour2000talking, wang2013students}. Additionally, increased retention has been linked to reformations in teaching and learning \cite{seymour2000talking,watkins2013retaining, griffith2010persistence}. Physics Education Research (PER) has also explored this topic finding similar results \cite{aiken2016methods,rodriguez2016gender,williams2017perc, harris2013perc,perkins2010perc}. This section summarizes the literature that has explored these themes.

A lack of interest, avoiding STEM courses, and other non-performance based measures can be indicators that students will switch from STEM \cite{seymour2000talking, chen2013stem, marra2012leaving}. \citet{seymour2000talking} cited loss of interest in STEM and an increased interest in non-STEM topics as a predictor of switching out of STEM. More recently, \citet{chen2013stem} examined STEM attrition rates and focused on students who leave STEM for non-STEM programs or those who dropped out entirely. \citet{chen2013stem} found that students who avoided STEM courses in their first year were likely to switch out of STEM. \citet{marra2012leaving} found that students who left engineering programs reported that a lack of belonging in engineering is a more important factor than performance related factors.

Teaching and learning reformations also have impacted students leaving and staying in STEM \cite{seymour2000talking, watkins2013retaining, griffith2010persistence}. \citet{seymour2000talking} found that poor teaching methods by STEM faculty influenced students to leave. The reverse has also been demonstrated: students who were exposed to interactive engagement techniques in introductory courses were more likely to persist in their STEM programs \cite{watkins2013retaining, griffith2010persistence}.  Student persistence in STEM has also been linked to attending colleges and universities that focus on teaching over research \cite{griffith2010persistence}.

Performance in coursework has also been highlighted as a contributing factor, such that better performance is tied to persistence in STEM \cite{griffith2010persistence,chen2013stem,shaw2010patterns,aiken2016methods, hackett1989exploration}. \citet{seymour2000talking} found a small but significant fraction of students who were enrolled as STEM majors but left STEM reporting conceptual difficulties with STEM coursework. Students required to enroll in lower math courses, and having poor grades in STEM courses were all indicators that students will leave STEM programs \cite{chen2013stem}.

In addition to performance at the university, the prior preparation of students has been demonstrated to effect outcomes at the university \cite{seymour2000talking, tai2001gender, sadler2001success, hazari2010connecting, schwartz2009depth}. \citet{seymour2000talking} found that many students believed their high school STEM education provided little to no preparation for the university. Reasons for this lack of preparation included that high school was too easy for these students, that ``gifted'' students often times were not taught study skills, that students experienced gender discrimination, and that the student's high school lacked resources. Teaching and learning in high school physics has also been linked to performance in university physics and STEM leaving \cite{seymour2000talking, tai2001gender, sadler2001success, schwartz2009depth}. Students who took high school physics courses that focused on ``deep and narrow coverage'' outperformed students who took ``broad and shallow'' courses in their university physics courses \cite{tai2001gender, sadler2001success, schwartz2009depth}. Ultimately, \citet{seymour2000talking} noted that students ``complain, with good reason, that they had no way to know how poorly they were prepared.'' Thus, high performance in certain high school contexts can be a predictor of leaving STEM in university.

Different demographic groups have a wide variety of experiences in their education that can impact their choice to stay or leave a major \cite{seymour2000talking, crawford1990gender, sadler2012stability, tyson2007science, tai2001gender, wang2013students}. Women and minority groups have reported a ``chilly environment" in the classroom and on-campus that is often less experienced by their white male counterparts \cite{crawford1990gender}; this remains true when students from different groups were similar in academic performance \cite{seymour2000talking}. \citet{seymour2000talking} found that curriculum pace, receiving poor grades while expecting high grades, and competition within STEM majors disproportionately affected men on their decision to leave a STEM major compared to women. \citet{seymour2000talking} attributed the ``weed-out process'' having ``a greater impact on young men because it carries messages which are intended to have meaning for them...[the weed-out processes] are obscure to young women, and they are thus less directly affected by them.'' Female students were also less likely to be interested in STEM in their senior year of high school in comparison to their freshman years while male counterparts' interest remained stable over that time \cite{sadler2012stability}. This is true even when they were enrolled in high level STEM courses in high school \cite{tyson2007science}. Female students have also been shown to earn lower grades in calculus-based physics courses when compared to male students with similar backgrounds \cite{tai2001gender}.

Students from racial and ethnic backgrounds that are underrepresented in STEM also have different experiences in the STEM programs, which can impact their choices surrounding traditional STEM degree programs \cite{seymour2000talking, tyson2007science, wang2013students}. These experiences begin pre-college where black and Hispanic students often have had less STEM opportunity than white students \cite{seymour2000talking, tyson2007science}. For example, black and Hispanic male students were less likely to have substantial STEM preparation in high school compared to white students; however, when these students had similar preparation, they pursued STEM careers in equal measure \cite{tyson2007science}. Additionally, \citet{seymour2000talking} found that under prepared students could appear ``over-confident'' in the STEM preparation because they excelled in less than average high school programs. Confronting that reality in college has lead to these students switching from STEM. Finally, participating in high school science and math courses generally has had a positive effect across all groups \cite{wang2013students}. However, students who identify as minorities are affected less positively by these programs than white and Asian students.

While a substantial amount of work has investigated students leaving  STEM, there has been less work that has focused on leaving physics, specifically. The work that has been done supports the broader work in STEM and has identified better performance \cite{aiken2016methods}, use of teaching reformations \cite{rodriguez2016gender, otero2006responsible, williams2017perc, harris2013perc}, and increased interest \cite{perkins2010perc} as all having had positive effects on the retention of physics majors. \citet{aiken2016methods} found that students' performance in introductory physics and calculus courses and when these students take these courses was important to completing a physics degree \cite{aiken2016methods}. In other work, physics majors were at least as likely to graduate with a physics degree who had been in reformed introductory courses as those taking traditionally taught courses \cite{rodriguez2016gender}. Interviews with physics majors who were also Learning Assistants \cite{otero2006responsible} have suggested that participating in a Learning Assistant program increased physics major retention \cite{williams2017perc, harris2013perc}. Students who become physics majors were also more likely to have expert-like beliefs as measured by the Colorado Learning Attitudes about Science Survey (CLASS) \cite{adams2006new} when they entered the university \cite{perkins2010perc}.

%% file: figure1.tex
\begin{figure*}[t]
\centering
\includegraphics[ trim={0 3cm 0 3cm}, width=1\textwidth]{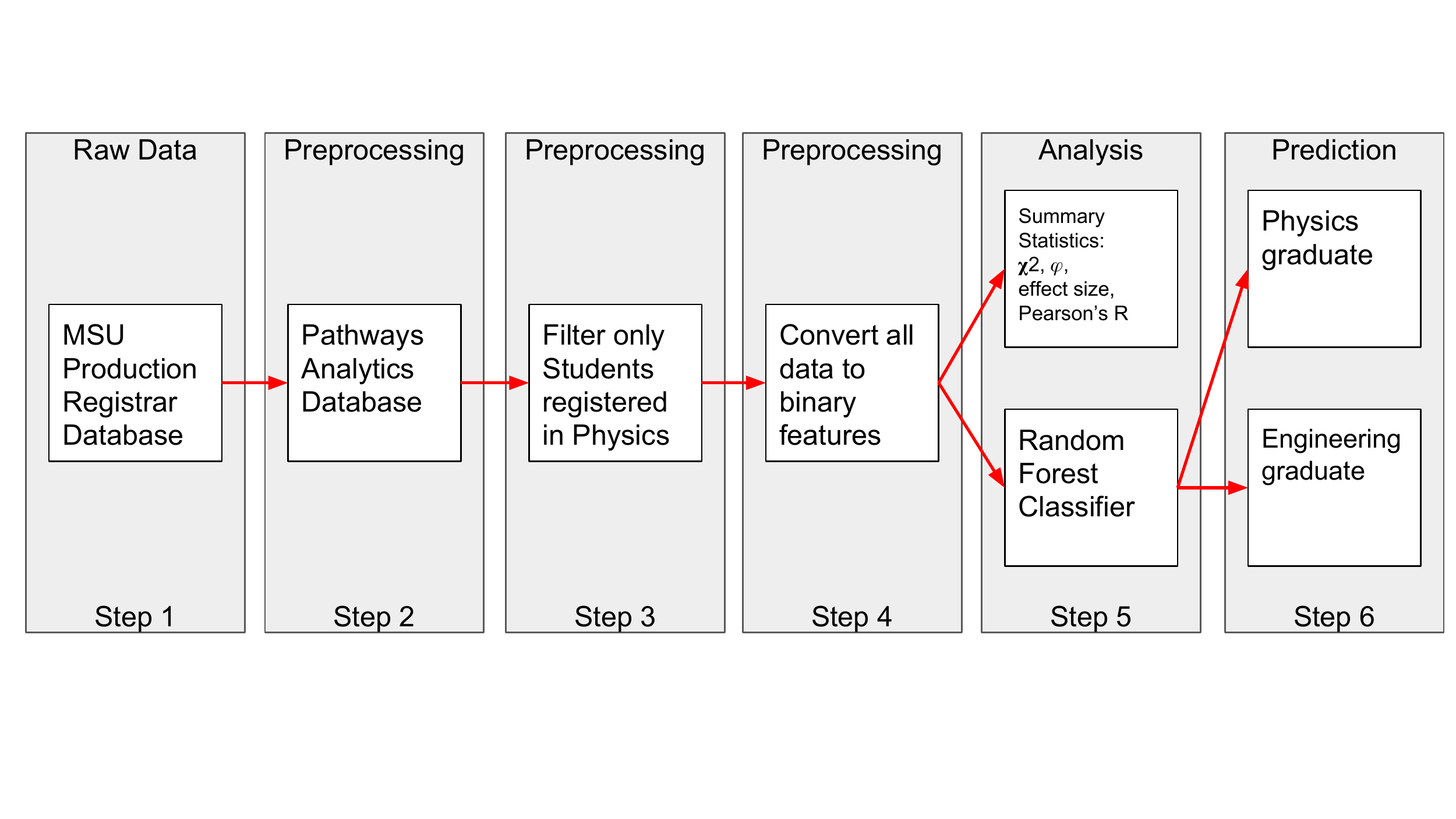}
\twocolumngrid
\caption{\label{fig:pipeline} Raw data is collected by the MSU Office of the Registrar and is used to build the Pathways Analytics Database. The data is filtered to only students who registered as physics majors and then further reduced to binary features. The analysis includes summary statistics (Table \ref{tab:contingency_table} and Figure \ref{fig:corrplot}) and the Random Forest Classifier predicting ``physics graduates" and ``engineering graduates" (Figures \ref{fig:featimportance_auc}, \ref{fig:aucacc}, \ref{fig:feat_over_time}).} 
\end{figure*}

%% file: data_based_on_figure.tex

This work focused on students enrolled in the physics bachelor's program who left for an engineering degree using data collected by the Office of the Registrar at Michigan State University (MSU). MSU is a large, land-grant, American university in the Midwest. It typically has had an enrollment of approximately 50,000 students and is predominately white (76.5\% across the entire university) \cite{msudescription}. University STEM degree programs are split between the College of Natural Science and the College of Engineering.

The data collected by the MSU Registrar included time-stamped course information that allowed for the examination of features related to students showing a lack of interest (e.g., students take courses much later than their first year). It also contained grade data for courses so performance metrics were examined. Throughout the study, teaching reformations were controlled for by examining the data set during a time (1993 to 2013) when there were no research-based teaching reformations enacted at MSU.

The data set used in the following analysis was built from database collected by the MSU Registrar. Prior to performing any analyses, the data was preprocessed to form what will be called the ``Pathways Analytics Database'' or ``Pathways database'', for short. The data was preprocessed following the pipeline shown in Figure \ref{fig:pipeline}, which is described below.

\subsection{\label{sec:pathways}Pathways Analytics Database}

The Pathways database contains three tables describing student demographics, time-stamped course data, and time-stamped major registration data (Step 2 in the data pipeline diagram in Figure \ref{fig:pipeline}). The Pathways database catalogs every course a student takes and every major they declare including their final degree. It includes grades for every course as well as transfer institution data for courses with transfer credit. It includes demographic information such gender and ethnicity. For some students, prior preparation data such as high school GPA and performance on the MSU math placement test is also available.

The data within the Pathways database was valid for students who began studying between 1993 and 2013.  The 2013 cutoff was used because students might not have had adequate time to reach graduation by Spring 2017 (the date when the data was pulled) if they enrolled past 2013. Prior to 1993, MSU was on a quarter system and there were major changes to degree programs and courses after the switch, thus, data prior to 1993 was not used in the analysis.

\subsection{\label{sec:filter}Filtering}

To investigate the research questions outlined in Sec.~\ref{sec:intro}, a set of appropriate filters was developed and applied to the Pathways database. With the focus toward investigating why students leave the physics major for an engineering degree, the data was filtered for only undergraduate students who at some point during their undergraduate program declared a major offered by the Department of Physics and Astronomy (Step 3 in the data pipeline diagram in Figure \ref{fig:pipeline}). Additionally, the data set was filtered to only students who ultimately received a degree from the either the Department of Physics and Astronomy or from the College of Engineering. The final data set did not include students who never completed a degree program nor did it include students who switched to non-STEM programs or other STEM programs. Students registered as physics majors who received degrees in physics or engineering made up 66.5\% of the students who ever registered as a major in the Department of Physics and Astronomy between 1993 and 2013. A total of 1422 students were analyzed in this study. In this work, the students who registered as physics students and then were awarded degrees from either the Department of Physics and Astronomy or the College of Engineering are referred to as either ``physics graduates'' or ``engineering graduates'', respectively.

The data for each of 1422 students was organized into single vector of features (or variables). These vectors contained all of the model features that were used to predict their final graduated degree. The features with summary statistics can be found in Table \ref{tab:contingency_table}.

Several features were included in the analysis based on previous literature described in Section \ref{sec:background}. For example, the demographic features were included because the experiences of female students and students who identify as ethnic or racial minorities have been shown to affect graduation outcomes in STEM \cite{seymour2000talking, chen2013stem, wang2013students}. In addition, grades in introductory physics and calculus courses were also included since, in prior work, course performance has been shown to be linked to students switching from physics programs to engineering programs \cite{aiken2016methods}.

Furthermore, the time when students take STEM courses has been shown to impact a student's success in a STEM program (earlier is better) \cite{chen2013stem}; thus, the time at which students take introductory physics and calculus courses have all been included in the analysis.

In addition to the features that have been discussed in prior literature, the impact of the individual courses on earning a degree in physics or engineering was also explored. Including such course-level registration features (specifically taking engineering courses or the first modern physics course) was based upon two hypotheses: (1) students who switched to engineering degree programs might have chosen to do so prior to actually switching in the MSU registrar database and thus may have signed up for required engineering courses, and (2) students who took the first modern physics course had invested in the physics degree, as this was the first course that was offered with only physics bachelor degree students being required to take it \cite{rodriguez2016gender}.

\subsection{\label{sec:binary}Converting to binary features}

For the analysis, all model features were converted to binary features (Step 4 in the data pipeline diagram in Figure \ref{fig:pipeline}). While some features were collected as binary (e.g., gender), some were converted to binary. An explanation for each converted feature appears below.

\textbf{Grades} - Grade features were reduced to a ``high/low'' binary feature and with a chosen cutoff at $\geq 3.5$. Grades can be considered ordinal data \cite{pike2008first}, and indeed were not on a strict interval scale (e.g. while the minimum grade point at MSU is 0.0 and the maximum is 4.0 there was no ability to earn a 0.5 grade point score). Grades in introductory physics and calculus courses rose beginning in the early 1990s. The average grade in physics and calculus rose by approximately 0.5 grade points (from $\sim$2.5 to $\sim$3.0 for physics, and $\sim$2.0 to $\sim$2.5 in calculus). Thus, the cut off of $\geq 3.5$ GPA was chosen because it was always above the grade increase.

\textbf{Transfer credit} - Students could come to MSU with transfer credit from Advanced Placement or from other institutions of higher education. When a student had a transfer credit for a particular course instead of a grade, the data was coded with a 0 in the high/low grade feature and a 1 in the transfer grade feature for the course.

\textbf{Time when courses were taken} - Time features were converted to on-time/late for each course. The split between on-time and late is different for first semester courses and second semester courses. For physics 1 and calculus 1, the cut off was set after the first semester. This was because the MSU physics department recommends that students take these courses in their first semester of enrollment. For physics 2 and calculus 2, the cut off was set after the first year of enrollment. This was because the MSU physics department recommends that students take these courses in their second semester of enrollment. In all cases, a positive response is when the student has taken the course prior to the established cut offs.

\textbf{Ethnicity} - MSU changed the way students reported their ethnicity over the course of the data set due to changes in the Integrated Postsecondary Education Data System (IPEDS) ethnicity definitions in 2007 \cite{ipedsdefinitions}. Thus, ethnicity definitions were first collapsed into the pre-2007 IPEDS definitions; they cannot be conversely expanded because the new definitions are more nuanced. The data were further collapsed into a binary feature indicating whether the student identified as white or Asian, or as a different reported ethnicity.

\textbf{Gender} - Gender was collected by the MSU Office of the Registrar as binary data (either male or female) and we used it as such in our model.

\textbf{Engineering Courses} - A feature was created that assessed whether or not a student had taken engineering courses. The data was restricted, per student, to the semesters when the student was registered as a \textcolor{ForestGreen}{physics major} in the Department of Physics and Astronomy. For example, a student who took one or more engineering courses during the semesters in which they were registered as a physics major was coded with a 1 for this feature.

\textbf{First Modern Physics Course} - A feature was created as to whether a student took the first physic course, modern physics, offered at MSU. This feature was coded with a 1 if a student was enrolled in this course at any point.

In addition, two model features which focused on prior preparation were also available for some students. These features included students' reported high school GPA and the score, if taken, that a student received on the MSU math placement exam. These two features require additional explanation that will appear in Section \ref{sec:results}. The model features described above form all of the features used in the various models throughout the study.




%% file: featuretable.tex
\begin{table*}[]
\caption{Summary statistics and contingency table analysis. The summary statistics are presented as percentages of the students who graduated with a degree (physics or engineering). Superscript ``a" denotes $p<0.05$, ``b" denotes $p<0.01$, and ``c" denotes $p<0.001$.}
\label{tab:contingency_table}
\begin{tabular}{l c c c c c}
		\hline
		\multirow{2}{*}{Feature} & Physics & Engineering & \multirow{2}{*}{$\chi^2$} & \multirow{2}{*}{$\phi$} & \multirow{2}{*}{\textit{V}} \\ 
		& Graduate (\%)& Graduate (\%)&&&\\\hline
		Took Modern Physics                           & 87.09          & 8.86               & 476.19$^c$       & 0.34         & 0.15       \\ 
		Took Engineering Course                       & 51.35          & 79.63              & 42.65$^c$        & 0.03         & 0.05       \\ 
		High Physics 1 Grade                          & 37.84          & 19.84              & 40.57$^c$        & 0.03         & 0.04       \\ 
		High Physics 2 Grade                          & 41.59          & 23.55              & 36.04$^c$       & 0.03         & 0.04       \\ 
		High Calculus 2 Grade                         & 25.98          & 12.30              & 35.39$^c$        & 0.03         & 0.04       \\
		Physics 1 On-Time                                & 69.22          & 46.96              & 30.58$^c$        & 0.02         & 0.04       \\ 
		Calculus 2 On-Time                               & 72.37          & 53.31              & 20.68$^c$        & 0.02         & 0.03       \\
		Female                                        & 16.67          & 10.71              & 9.29$^b$         & 0.01         & 0.02       \\ 
		Calculus 1 Transfer Credit                    & 55.56          & 44.58              & 8.58$^b$         & 0.01         & 0.02       \\
		Physics 2 Transfer Credit                     & 20.72          & 14.82              & 7.03$^b$         & 0.01         & 0.02       \\
		Calculus 1 On-Time                               & 84.08          & 73.28              & 5.28$^a$         & 0.00         & 0.02       \\
		Physics 2 On-Time                                & 31.83          & 26.46              & 3.53          & 0.00         & 0.01     \\ 
		Physics 1 Transfer Credit      	              & 29.73          & 25.27              & 2.58         & 0.00         & 0.01       \\
		White or Asian & 86.64          & 82.14              & 0.85         & 0.00         & 0.01       \\
		High Calculus 1 Grade                         & 16.82          & 15.08              & 0.67         & 0.00         & 0.01       \\ 
		Calculus 2 Transfer Credit                    & 32.88          & 33.33              & 0.02         & 0.00         & 0.00    \\\hline
		Total (\textit{N}) & 666 & 756 &&&\\\hline
	\end{tabular}
\end{table*}

%% file: methods.tex
In this work, contingency tables \cite{everitt1992analysis} and binary classification models \cite{friedman2001elements}, specifically Random Forests, were analyzed to predict whether a student would receive a degree from the Department of Physics and Astronomy or switch to an engineering degree (Step 5 in Figure \ref{fig:pipeline}). Contingency tables have been widely used in PER (e.g., \citet{finkelstein2005learning}). Often, researchers employ the $\chi^2$ statistic and effect sizes to assess the correlation of frequency data. For all features described above $\chi^2$ statistics and Cramer's \textit{V} effect sizes \cite{acock1979measure} were calculated (Table \ref{tab:contingency_table}). Additionally the Pearson correlation coefficients \cite{lee1988thirteen} were also calculated (Figure \ref{fig:corrplot}).

Binary classification models, specifically logistic regression, have also been used in physics education research (e.g., \citet{dabney2014comparative}). In general, a binary classification model predicts an outcome that is discrete and (usually) binary. For example, logistic regression can be used to predict if a student stays or leaves STEM, earns a grade above or below a certain level, or whether or not a student completes a specific course. This technique can be extended to a multi-class outcome variable; however, for the purpose of our analysis, the outcome variable of interest (physics graduate vs. engineering graduate) was binary. 

Binary classification models are a supervised learning technique that can consist of various algorithms (e.g. logistic regression). In this paper, we have used the Random Forest algorithm for classification \cite{breiman2001random} because the underlying features in our analysis were binary (Section \ref{sec:binary}). As it is a fairly new method to PER, this section will provide a summary of the Random Forest algorithm \cite{breiman2001random}. Additionally, it will introduce how to assess the model predictions and associated inferential statistics.For a thorough review of the Random Forest algorithm, see Young \textit{et al.} \cite{young2018using}.

\subsection{Random Forest}

Random Forest classifiers use the mode output of a collection of decision trees to predict if the input data belongs in one class or another \cite{breiman2001random}. The model produces ``feature importances'' which represent the average change in the decision criterion for each feature in the data set. The larger the feature importance, the more important the feature is to the model. It is important to note that feature importance is measured \textit{relative} to the other features in the model. An exhaustive review of Random Forest algorithms and their uses can be found in references \cite{breiman2001random, friedman2001elements, geron2017hands}.

The power that the Random Forest algorithm has over a more traditional model (e.g., logistic regression) is that it is an ensemble model \cite{breiman2001random}. Ensemble models are a collection of sub-models whose outputs, taken together, form the prediction. Because Random Forest is a collection of decision trees, the Random Forest can fit multiple sub-populations within a data set. Thus the ensemble of trees in the random forest can fit subsets of the data without overly biasing the model output \cite{svetnik2003random}. By contrast, a model like logistic regression attempts to fit a hyperplane to the entire input data set \cite{hosmer2013applied}. In a Random Forest, no one tree in the forest is required to be predictive of all the data. Rather, it is the collection of decision trees that make up the Random Forest that form the predictions. 

\subsection{\label{sec:eval}Evaluating a classification model}

In PER, researcher are often concerned with not only providing predictive outcomes, but explaining the system that being studied. However, analyzing the predictive output of an explanatory model is important even if it is not the intent \cite{shmueli2010explain}. In doing so, studies can provide a basis for comparison with other models that may also fit data from different settings. Below, we introduce how to evaluate the predictive output of a classification model. These methods are applicable to all classification models (e.g., logistic regression), not only the random forest model used in this paper.

Model predictions are evaluated in a number of ways such as accuracy, receiver operating characteristic (ROC) curves, and the area under the ROC curve (AUC) \cite{friedman2001elements}; the best model can be found using a grid search \cite{geron2017hands}. For this work, the best model was defined as the model with the highest combination of metrics that produced good predictions. Below, we discuss these evaluations metrics and the relationships between them.

Accuracy is the ratio of true predictions to all predictions made. This measurement has a caveat; when the data is class imbalanced (i.e., classes are not split evenly \cite{chawla2004special}), the accuracy can be less meaningful \cite{geron2017hands}. For example, if 90\% of the data belongs to class A, having a prediction algorithm that assigns class A to all data produces a 90\% accuracy. In this analysis, the data does not have a large class imbalance (see Section \ref{sec:lopsided}).

Receiver Operator Characteristic (ROC) curves compare the true positive rate (TPR) and the false positive rate (FPR) for a variety of cutoffs \cite{fawcett2004roc}. The TPR is the ratio of true positives to the sum of true positives and false negatives. The FPR is 1 minus the ratio of true negatives to the sum of true negatives and false positives.

ROC Curve plots have a certain geography to them. The diagonal serves as a boundary; the model is better than guessing if the curve is above the diagonal and worse than guessing if the curve is below it. A curve trending towards the upper left is one that is approaching perfect classification \cite{young2018using}.

The area under the ROC curve (AUC) \cite{hanley1982meaning} provides an additional summary statistic for interpreting the quality of a classification model. The closer the AUC is to 1, the better the performance of the model. Performance in this case is defined as the ratio of TPR to FPR.

Random Forest models have a number of parameters that govern the size and shape of the forest (e.g., number of decision trees, number of features allowed per tree, the decision criterion, etc.). Because different parameter choices can produce models with different qualities of predictions, we employed a grid search to determine the best choice of parameters for our models \cite{hsu2003practical}. A grid search is a method used to test the total combination of a range of possible model parameters to return the highest fitting scores. In this work, we assessed the model's performance using accuracy and AUC for classifiers. The range of parameters and the Python scripts used to test and evaluate the models in this paper can be found in the online Jupyter notebook \citep{gitrepo}.

\subsection{\label{sec:lopsided}Training and testing the model}

Testing a model's predictive power requires splitting the data into a training and a testing data set; this worked used a ratio of 70:30. That is, 70\% of the data was used to train the model and 30\% of the data was used to test the model's predictions. This 70:30 ratio is a common choice for many machine learning techniques \cite{geron2017hands}. The data is randomly sampled into a testing and a training set without replacement. The testing data is sometimes called "hold out data" because it is held from the model and is not used for fitting the data \cite{geron2017hands}. In this work, the ratios between the two graduating outcomes for the training and the testing data were the following:

\begin{itemize}
    \item \textbf{Physics} Training: 45.7\%, Testing: 47.3\%
    \item \textbf{Engineering} Training: 54.3\%, Testing: 52.7\%
\end{itemize}

Thus within the data, the graduating degrees were roughly equivalent in shape and there was little class imbalance.

To produce the highest AUC and accuracy, the model was run repeatedly through parameter variations via a grid search. By performing a grid search, 1920 parameter combinations were tested. The parameter ranges were chosen vary around the default values of the parameters as specified in the sci-kit learn documentation for random forest classifiers \cite{pedregosa2011scikit}. For example, the maximum number of features available to a tree uses the minimum of the default: $n_{features}=\sqrt{n_{features,max}}$ up to half the maximum. The maximum number of feature combinations would exceed the computational capacity available for this project thus some limits were placed on the grid search (such as not varying $n_{features}$ between the true minimum and maximum number of available features) to minimize computational time. The final output model parameters were then used to fit the Random Forest classifier to the training data set. The various parameters found in the Jupyter notebook mentioned above \citep{gitrepo} include commonly varied parameters such as the number of trees in the forest and the depth of the forest.

Because of the observed grade increase over time (described in Section \ref{sec:binary}, grades in introductory courses have some time dependence. Thus, it was expected that other features might be time dependent as well. To investigate this and how it might effect the analysis, a sliding window approach was used to analyze this time dependence \cite{dietterich2002machine}. To explore if our model and the resulting features were temporally invariant, the data was split into 17 windows centered around each year from 1996 to 2013. For example, a 4 year window centered on 1996 included data from 1994 to 1998. The model was then refit on each individual window of data. The feature importances were compared for each time window to investigate which features remained invariant over time. To assess the quality of the window size, we used 4, 6, and 10 year window sizes from the beginning of the data set (1996) to the end (2013) in increments of one year. Since each window size produced similar results, only the 6 year window is reported in this paper. 

\subsection{\label{sec:reduce}Reducing the model to the minimum viable features}

The final way that a model can be evaluated is combining all of the previously mentioned evaluation methods (accuracy and AUC) to determine what is the minimum subset of features necessary to construct a viable model. This method is known as ``recursive feature elimination'' \cite{guyon2002gene, guyon2006introduction}. Using the best parameters provided by the grid search, the recursive feature elimination (RFE) algorithm begins with all of the features available, and removes the least important features (based on feature importance magnitude) one by one until only the most important feature is left. This process produces AUC and accuracy curves that can be compared to one another to determine the most predictive model with the least number of features. In a sense, this process is similar to finding the fewest number of features in a linear regression model that produces a fit that is not statistically different from the model with all the features. In RFE, features that are less important to model prediction have minimal effects on AUC or accuracy when included in the model. It is then assumed that such features do not represent an aspect of the data that is important for prediction.

\section{\label{sec:allthemodels}Analyzed Models}

Throughout the analysis, multiple models were investigated. Below we describe the details of each model that was used:

\begin{enumerate}
    \item \textbf{Main Model}: The main model included all available features and was used to perform a grid search to find the best parameters that defined the forest (Figure \ref{fig:featimportance_auc}). This model used the entire data set (N=1422 students).
    \item \textbf{Recursive Model}: The recursive feature elimination (RFE) produced as many models as there were features and used the best parameters to define the forest from the grid search in the Main Model. This was used to describe which features impacted the model the most (Figure \ref{fig:aucacc}). This model used the entire data set (N=1422 students).
    \item \textbf{Sliding Time Window Model}: The sliding time window model, produced a model for each time window using the best parameters defined by the grid search in the Main Model. It used subsets of data built from each time window for the training and the testing data (Figure \ref{fig:feat_over_time}). This model used the entire data set (N=1422 students).

    \item \textbf{Prior Preparation Model}: A final model was built from training data that had been reduced using complete case analysis. Complete case analysis excludes all data that has missing data. In this case, high school GPA and the MSU math placement test score were added to the model. The model was then trained using the complete case for high school GPA or the MSU math placement test score to determine if these features were important predictors to students staying in physics. This model also used the parameters from the grid search in the Main Model. This model used two subsets of the data set (N students with HS GPA = 1037 (73\%), N students with math placement score = 700 (49\%)).
\end{enumerate}

%% file: figure2.tex
\begin{figure}[t]
\centering
\includegraphics[width=250px]{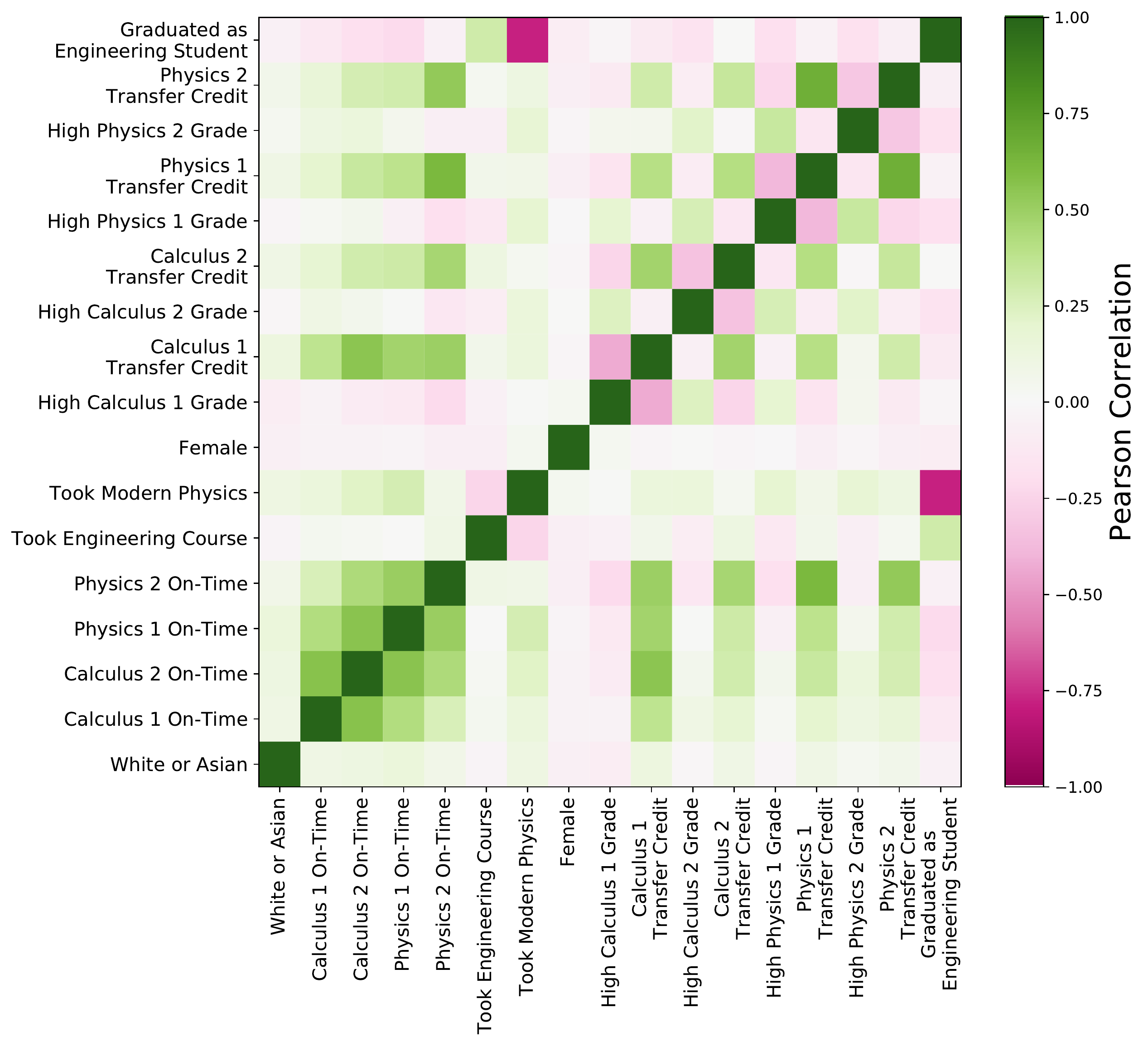}
\caption{\label{fig:corrplot}Pearson correlation matrix for features in the Main Model.}
\end{figure}

%% file: figure3.tex
\begin{figure*}[t]
\centering
\includegraphics[width=1\textwidth]{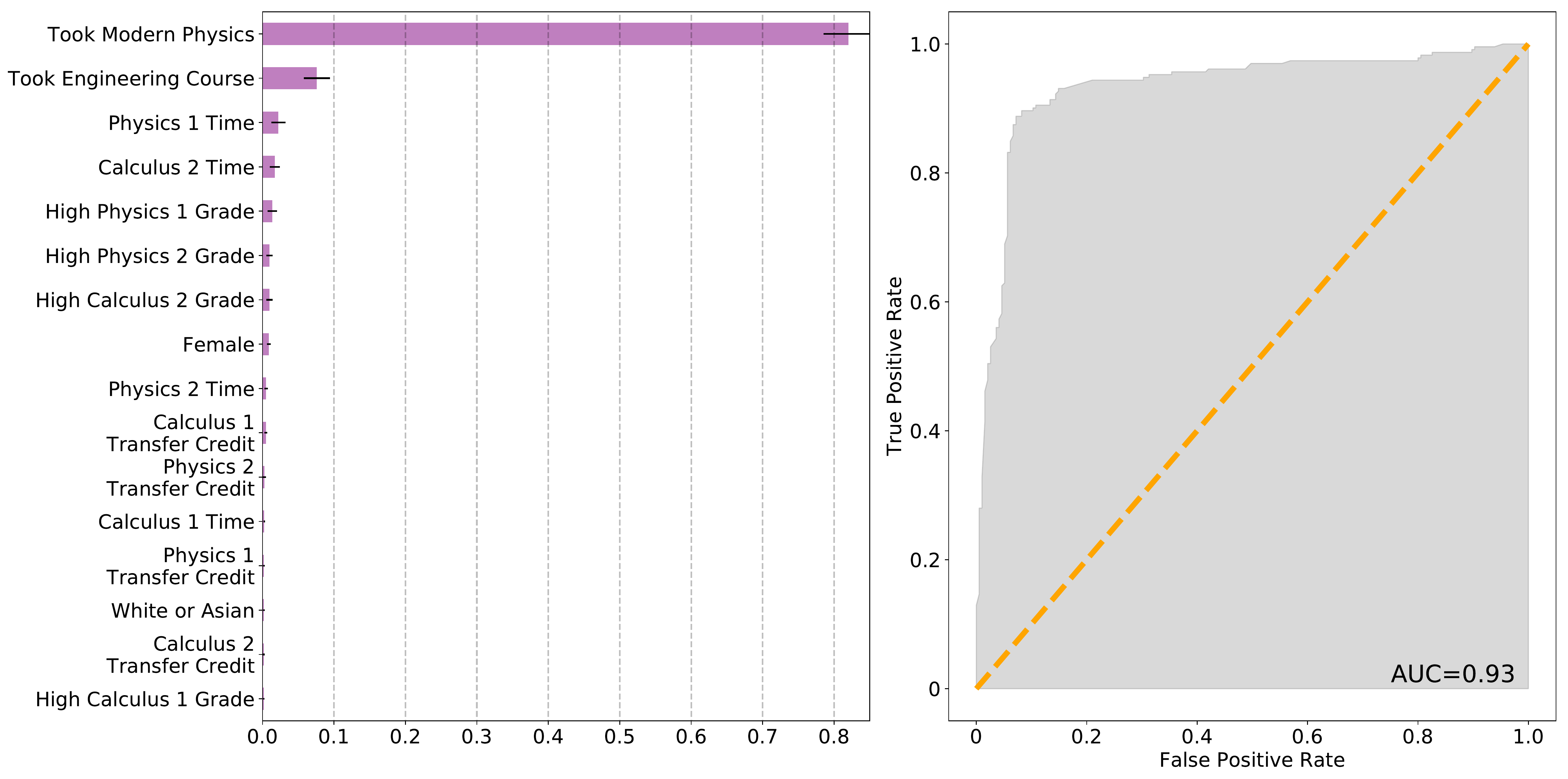}
\put(-418,30){(A)}
\put(-175,30){(B)}
\caption{\label{fig:featimportance_auc} (A) Feature importances for the Main Model. The error bars represent the standard error for the distribution of feature importances across all trees within the forest. (B) The ROC curve for the Main Model. The dashed line represents random chance.}
\end{figure*}

%% file: figure4.tex
\begin{figure}[t]
\centering
\includegraphics[width=235px]{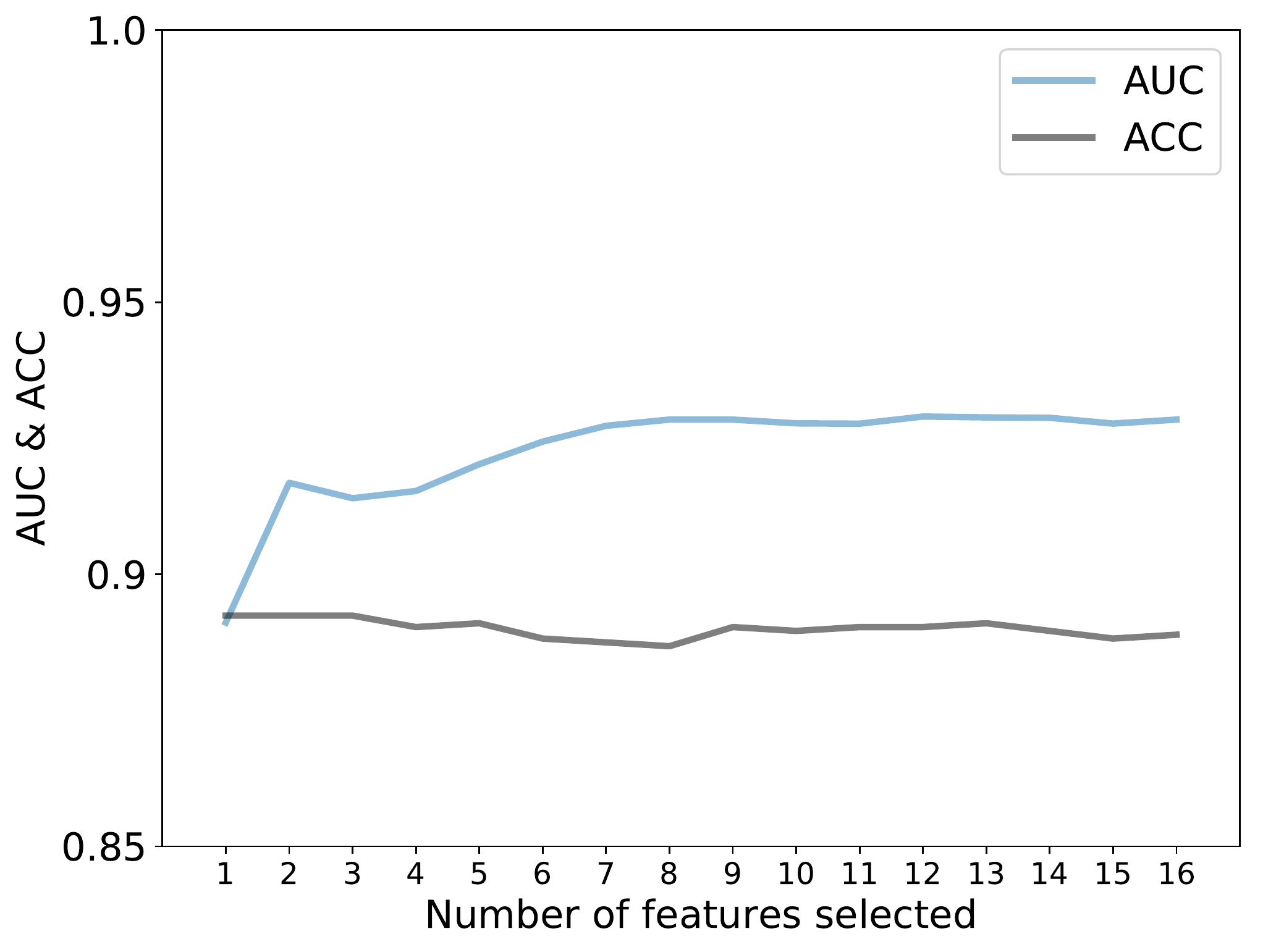}
\caption{\label{fig:aucacc} The AUC and ACC for each number of features selected. The features are ordered the same as Figure \ref{fig:featimportance_auc}.}
\end{figure}

%% file: figure5.tex
\begin{figure}[t]
\centering
\includegraphics[width=235px]{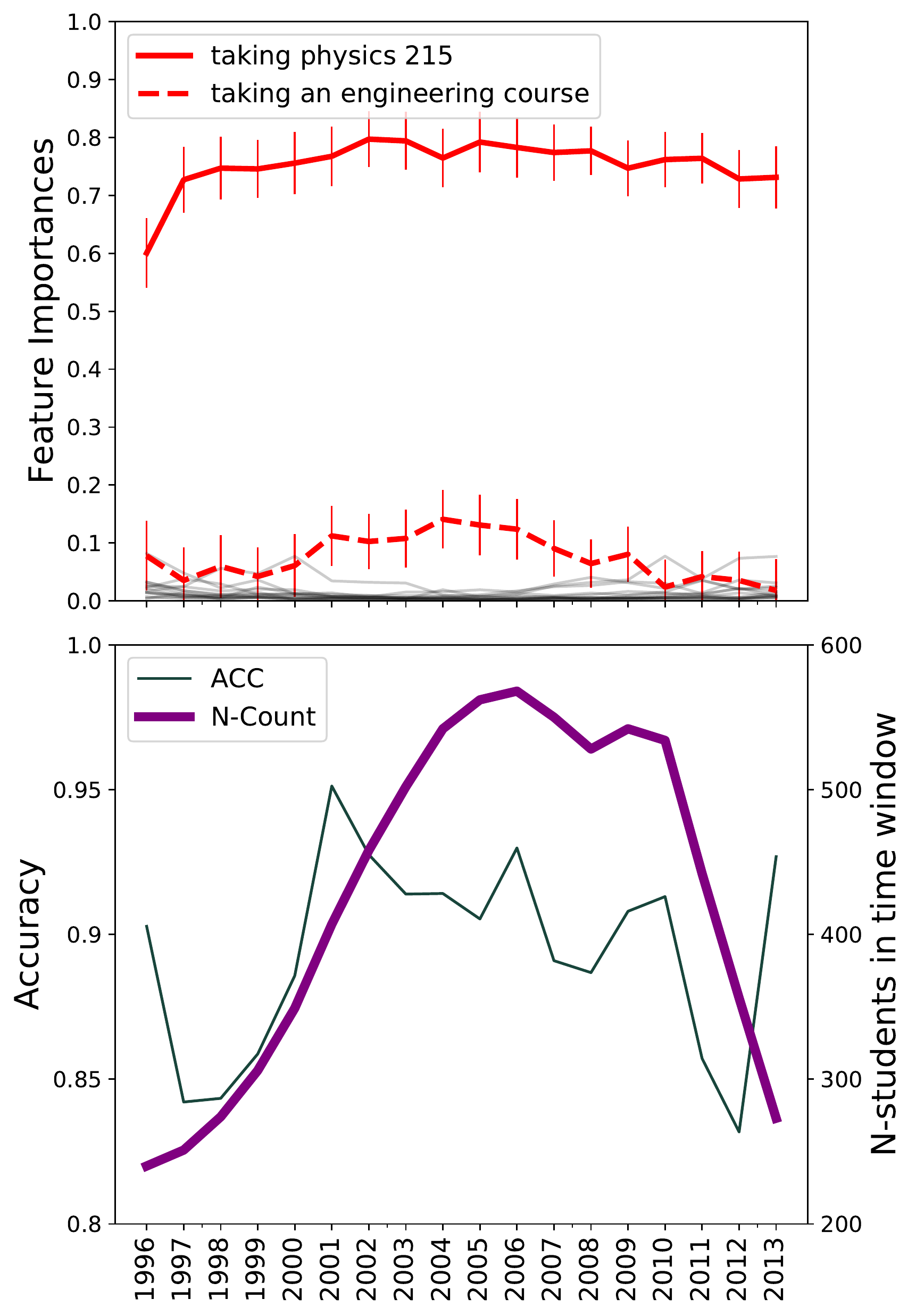}
\put(-50,315){(A)}
\put(-50,155){(B)}
\caption{\label{fig:feat_over_time} Sliding Time Window Model. (A) The original grid searched cross validated model parameters were fit to data from sliding time windows ($\pm 3$ years) with the window center starting in 1996 and running through 2013. The error bars represent the standard error of the feature importance. (B) The Accuracy (ACC) for each of the time windowed models. The dip in the number of students in later years was due to the sampling method; since no students who begin after 2013 are included in the data set, the total number of students from 2010 begins to decrease.}
\end{figure}

%% file: results.tex
A contingency table analysis demonstrated that taking modern physics had the highest effect ($\phi_{modern}=0.34$, $p<0.001$) on students who earned a degree from the Department of Physics and Astronomy (Table \ref{tab:contingency_table}).
While this feature held the highest effect size, it is still relatively small \cite{ferguson2009effect}.
All other features demonstrated negligible effect sizes. Previously, it was demonstrated that high grades in physics and calculus might indicate students will remain in a physics degree program \cite{aiken2016methods}; however, through this analysis it was found that high grades in both courses had negligible effect sizes on earning a degree from the Department of Physics and Astronomy. Outside of taking modern physics and the prediction feature (graduating with a physics or engineering degree), features in this data set showed low linear correlations with each other, but there were still interesting structures (Figure \ref{fig:corrplot}). For example, taking the modern physics course showed a low anti-correlation with taking an engineering course while registered as a physics major (Pearson's R=\phystakingeng). Results also showed that there was a positive correlation between the times at which students take courses. This is likely due to these courses having pre-requisites from one to the other. In addition, there were small anti-correlations between high grades and transfer grades. This does not indicate that a student with transfer credits for a course might receive low grades for the same course. Given that transfer grades and high grades were mutually exclusive, a student could not have a high grade from an MSU course and have transferred in credit for that course; this can be explained by how we constructed our data set. Contingency tables and Pearson correlations can give some insight into the data we gathered; however, they cannot give predictions about more individualized results or provide general inference about physics students who switch to engineering. To further explore this, a Random Forest classifier was employed.

A Random Forest classifier (the Main Model from Section \ref{sec:allthemodels}) was built to predict whether a student would graduate with a physics or engineering degree based on the features in Table \ref{tab:contingency_table}. The model demonstrated that taking modern physics was of greatest importance to the model's prediction (Figure \ref{fig:featimportance_auc}). This model's predictive ability is high demonsrated by (AUC=0.93) the ROC curve being well above the random guessing discrimination line.  No other feature had an importance above 0.1. The calculated feature importances are a measure of the mean decrease in Gini impurity each time the feature was used in a tree \cite{breiman2001random, pedregosa2011scikit}. The second most important feature was a student taking an engineering course while registered as a physics major. Using recursive feature elimination (Recursive Model in Section \ref{sec:allthemodels}) and comparing the AUC and accuracy, the optimum model had two features: taking modern physics and taking an engineering course as a physics major (Figure \ref{fig:aucacc}). Including any additional features reduced the accuracy of the model. Figure \ref{fig:aucacc} also shows that the AUC increased at a small expense of accuracy by including additional features: having a high grade in physics 1, physics 2, and calculus 2, taking calculus 2 in the first year, and whether or not the student is female. Including features beyond this did not increase AUC or cause noticeable changes in accuracy.

Given the observed increase in grades in introductory physics and calculus courses over time, the time-dependent nature of our findings were investigated. A Random Forest classifier trained on sliding windows of data subsets (the Sliding Time Window Model in Section \ref{sec:allthemodels}) also demonstrated that taking a modern physics course was the most important predictor for a student to receive a physics degree (Figure \ref{fig:feat_over_time}) at any point in time (i.e., for every window scale centered on any given year). There was little variation in the model for other features outside of the slight increase and decline in the feature importance for taking an engineering course while registered as a physics major.

MSU uses a custom math placement test to place students in math courses who do not have AP credit or high SAT/ACT math scores. This data was not present for all students as they might have had a high school transcript that resulted in them not having to take such assessments. Thus, student performance on this test and a reported high school GPA existed for subset of the data. 1037 students (73\%) had a reported high school GPA; 700 students (49\%) had a reported math placement score; and 604 students (43\%) had both reported. These features were not pre-processed like the other features in the data set and thus, represent the actual high school GPA reported or the score the student received on the math placement exam. These features were used in the Prior Preparation Model (see Section \ref{sec:allthemodels}). These features demonstrated no increase in the overall predictive power of the models for this subset of data. Because these features had no impact on model prediction, and they were missing for a large fraction of the data, these prior preparation features were excluded from all other models. Furthermore, because these features did not impact model prediction, they were not imputed.

%% file: discussion.tex
This work demonstrated that taking the first course in modern physics was the single most important feature for predicting if a student earned a bachelor's degree from the Department of Physics and Astronomy at Michigan State University. It additionally demonstrated that taking an engineering course while registered as a physics major was an indicator that students will switch and eventually earn a degree from the College of Engineering. Further, it was found that performance as measured by grades in introductory physics and calculus could have small effects on whether a student earned a physics degree or not.

We set out to answer three questions:

\begin{enumerate}
  \item Which of the factors identified in the literature impact students to remain a physics major or leave the physics bachelor's program for an engineering degree?
  \item What were the effect sizes of these performance factors and how do they impact models that compare the effect of various factors against each other?
  \item Through this analysis, What did we learn about students who register for physics but leave for engineering (i.e. intra-STEM switchers)?
\end{enumerate}

Section \ref{sec:background} describes multiple reasons why students might leave STEM. These include a lack of interest, poor teaching in STEM courses, performance in university STEM courses and prior preparation, and differential experiences for different demographic groups. For most of the time period of this study, there were no systematic research-based changes to teaching practices at MSU. Thus, from this data, it is not possible to provide commentary on how poor teaching and how course transformations or changes to pedagogy might have affected student retention in the physics major. Future work will examine how current course transformations in the Department of Physics and Astronomy at MSU (e.g., \cite{irving2017p3}) have changed physics major recruitment and retention. 

Below, we discuss the unique results from our analysis, namely, the role of the third-semester, modern physics course and how taking engineering courses as a physics major play in earning a physics degrees. Then, we discuss the roles that previous observed features (ie., lack of interest and performance in STEM courses) played in our work. In future sections, we will discuss the limitations and implications to this research study (Section \ref{sec:limitations}) along with the affordances that the model prediction assessment has above more traditional modeling approaches (Section \ref{sec:methodsdiscussion}).

\subsection{The importance of the first modern physics course}

We found that the most important feature in our model that predicted which degree a student earned was whether or not a student took the modern physics course (Figure \ref{fig:featimportance_auc}A). This result remained important as other features were eliminated (Figure \ref{fig:aucacc}) and was consistent over time (Figure \ref{fig:feat_over_time}). Finding that taking modern physics is the most predictive feature in our model bolsters the assumptions made in prior work. In \citet{rodriguez2016gender}, the authors decided to filter their data based on whether or not students had taken their modern physics course. If a student had not taken this course, \citet{rodriguez2016gender} did not consider the student to be a ``physics major'' in their data set. Our work further supports this assumption of taking the first required physics course for all physics majors (i.e. modern physics) was the most predictive feature in the data for finishing with a physics degree through all analyses.

The nature and role of this course in MSU's Department of Physics and Astronomy suggests that such a course should be a strong predictor of completing the physics degree program. The Department expects that students will take modern physics in their third semester of enrollment. It is the student's first exposure to thermodynamics and quantum mechanics at MSU and is also the first physics course that is not required for any major outside of Department of Physics and Astronomy. The course requires students to have completed the introductory physics sequence and be, at a minimum, concurrently enrolled in a multivariate calculus course.

Because this course predicted whether a student stayed in physics or left for an engineering degree, it could serve as a good outcome variable for future analyses. That is, future work will investigate which features predict if a student is likely to take the modern physics course. Understanding which features impact whether or not a student will take this course can help departments understand the potential pathways for future physics majors.

\subsection{Leaving for Engineering}
Taking an engineering course while registered as a physics major was a smaller, but still predictive, feature in the models. This could indicate that students who leave physics for engineering might not have ever intended to stay in physics, or, perhaps, that they might have found the applied nature of engineering more attractive. Furthermore, the experiences that students encounter in early introductory physics courses could of driven them away from physics and ultimately graduating with an engineering degree.

In contrast to students that leave the physics program for other majors \cite{aiken2016methods}, students at MSU who leave the physics for an engineering degree rarely take physics courses beyond the two-semester introductory sequence. Thus, taking an engineering course while registered as a physics major might be a relatively strong signal that a student intends on leaving physics for engineering. It could also be a signal that the student plans to dual major (something our models do not account for). These explanations cannot be confirmed by the data and approaches used in this study and will be further explored in future work. The precise reasons underlying intra-STEM switching are better unpacked using qualitative approaches.

Through the sliding window analysis, results suggested that students taking engineering courses while registered as physics majors had its largest feature importance in the mid 2000s (Figure \ref{fig:feat_over_time}). This was due, in part, to a sharp visual, but not statistically significant ($\chi^2=2.66$, $p=0.26$), increase in students switching from physics to engineering between 2005 and 2011 (see the Jupyter notebook \cite{gitrepo} for supplemental figures). Through discussion with relevant MSU faculty, we could find no credibly documented reason why these students might have left the Department of Physics and Astronomy for the College of Engineering in higher numbers during this period.

Though the more traditional, population-level analysis showed that the effect size of taking an engineering course was negligible ($\phi_{engineering} = 0.05$, $p<0.001$, see Table \ref{tab:contingency_table}), our Random Forest model did see an increase in the predictive power via the AUC when this feature was included (see Figure \ref{fig:aucacc}). However, ultimately, taking an engineering course had a negligible effect size. This implies that subtle features in our data set might not be best explained by population statistics (e.g., $\chi^2$, effect size). This observation should encourage researchers to use both descriptive and inferential analysis to conduct studies.

\subsection{Lack of interest}

While not measured directly, three of the features served as proxies for interest in earning a physics degree:

\begin{enumerate}
    \item The time when students took introductory physics or calculus courses relative to their enrollment at MSU (later may indicate less interest),
    \item Whether or not a student took an engineering course while registered as a physics major (doing so may indicate less interest), and
    \item If a student had transfer credit for physics courses (having credit may indicate greater interest).
\end{enumerate}

\citet{aiken2016methods} demonstrated that students who switched from physics to engineering were likely to enroll in introductory physics courses later than those who stay in physics; also confirmed in Table \ref{tab:contingency_table}). In the Main Model, the 3rd most important feature measured whether students take physics 1 in their first semester (Figure \ref{fig:featimportance_auc}). However, ultimately the feature has such little importance that it does not show a large change in AUC scores when added to the model (Figure \ref{fig:aucacc}). Thus, while it is true that engineering graduates take introductory physics courses later in their academic career, it does not seem to have a large impact on whether a student earns a physics or engineering degree relative to the other features in the model. When compared to the \citet{chen2013stem} result, namely, if students avoid STEM courses in their first year that is indicative of leaving STEM, the results presented above demonstrated that perhaps when students take their physics courses was inconsequential to switching from physics to engineering. However, this could be a signal that only certain STEM courses are indicative of moving within STEM. \citet{chen2013stem} highlights mathematics, saying ``proportionally more STEM leavers than STEM persisters did not earn any math credits in their first year.'' In this study, this effect was not observed as students are required to complete the introductory calculus sequence to earn a degree in either physics or engineering.

Ultimately, 613 (81\%) students who left for engineering took at least one engineering course while registered as a physics major. These courses are likely to be prerequisites to switch majors. For example, the most likely engineering course to be taken is CSE 231, Introduction to Programming 1; 326 (43.1\%) students switching from a physics degree to an engineering degree took this course. It is a prerequisite for admission into the College of Engineering to pursue a degree in Computer Engineering, Computer Science, and Mechanical Engineering \cite{engadmin}. Additionally, \citet{aiken2016methods} found that students who switched to engineering were more likely to take introductory physics in their second and third semesters whereas students who stayed in the physics program were more likely to take physics in their first semester. Taking introductory physics in the second and third semester is explicitly recommended by the Mechanical Engineering department at MSU \cite{engphysicsdates}. Thus, the courses that students take and when they choose to take them are indicators of a lack of interest in physics and therefore, will ultimately leave the physics major for an engineering degree.

Finally, the engineering graduates also had fewer transfer credits for physics courses than physics degree earners. Students who switched to engineering from physics were also less likely to take upper division physics courses in comparison to the students who switched to other majors from physics \cite{aiken2016methods}. These findings suggest that students who switched to engineering demonstrate less interest in physics based on what courses they choose to take early in their college experience, including their pre-MSU academic careers. 

It is important to note that the above claims about students' interest in physics are made from features that may be considered ``measures of interest''; these features are only proxies for a student's expressed interest in physics. Further work should be done to explore these claims explicitly since this is outside of the scope of this paper.

\subsection{\label{sec:performance}Performance in coursework}

A result of the above analysis that is in tension with prior work performed by \citet{aiken2016methods} is the small impact that grades have in predicting whether a student will earn a physics or engineering degree \cite{aiken2016methods}. Previously, researchers found that students who switched from physics to engineering performed below average in introductory physics and calculus courses in comparison to students who stayed in physics \cite{aiken2016methods}. It was assumed in the prior study that performance in a course could have a profound impact on a student's persistence in the physics major. Moreover, it has been well-documented that such performance measures are important to STEM persistence \cite{seymour2000talking}. 

In the current work, course performance, measured by grades, was significant but with a negligible effect size in the population level analysis (Table \ref{tab:contingency_table} and Figure \ref{fig:corrplot}). However, none of the Random Forest models found these performance features to be particularly important (measured by feature importance) to make accurate predictions. We posit three reasons for this finding. First, the methods used in the previous study \cite{aiken2016methods} were different than those used in the analysis described above. Here, using Random Forests, models were constructed to compare the relative effect of the explored features. In \citet{aiken2016methods}, only the effects of individual features were explored. Second, in previous work, grades were treated as a continuous feature and in the above analysis, they are treated as ordinal features. Finally, it might be that grades were not an important indicator for students leaving one STEM program for another STEM program (i.e., leaving physics for engineering).

In \citet{aiken2016methods}, grades were compared using Z-scores \cite{kreyszig2010advanced}. The Z-score normalizes all data to unit variance; thus, the larger an outlier, the more weight (visually) can be given to the score. As the goal of prior work was to provide methods for visualizing these types of data, the decision to use Z-scores was driven by considering compelling ways to represent our data, but the resulting analyses might not have been appropriate. Additionally, the previous work did not make any comment on an ``effect size'' that could represent the population comparisons effectively. While \citet{aiken2016methods} state that physics graduates performed better than engineering graduates in their introductory physics and calculus courses, the degree to which they performed better was not well quantified. This could be because the analyzed Z-score did not provide a proper normalization given that the underlying grade data is ordinal \cite{fielding2003multilevel}.

In the present case, conflicting results were found. At the population level, having a high score in physics 1, physics 2, and calculus 2 was statistically significant. But, each feature had a negligible effect on graduating with a physics degree (Table \ref{tab:contingency_table}, Figure \ref{fig:corrplot}). However at the individual level, (i.e., the Random Forest classification modeling) grades were not important features (Figures \ref{fig:featimportance_auc} \& \ref{fig:feat_over_time}). In fact, through recursive feature elimination, performance in these courses did not substantially impact the reduced models (Figure \ref{fig:featimportance_auc}).

By comparing the descriptive analysis (using population statistics) to the inferential analysis (using the Random Forest model), conflicting results were found, which indicate how the two different approaches can compare the relative impact of different features. This conflict might simply be due to the differences one encounters when conducting analysis using descriptive statistics compared to modeling; the descriptive population statistics aggregate all of the available data into a single result. In the descriptive analysis, a comparison was made between the two groups (physics and engineering graduates) and because the number of student records was reasonably high, we might expect to find a statistically significant difference in these distributions. The Random Forest model, however, places each student record into a class (physics or engineering) depending on the available data and computes the \textit{relative} importance of each feature against the others.

Grades might also not have been important for students leaving physics programs for engineering programs. \citet{seymour2000talking} identified the assumption of grade differences as a ``barrier'' to understanding why ``able students'' leave STEM. In the above study, students who left physics for engineering are ``able'' in Seymour's words. Course grades were included in the above study because \citet{aiken2016methods} demonstrated that they might be important and that initial findings demanded further exploration. However, in this analysis, grades were shown to have no effect on the Random Forest models and negligible effect sizes. \citet{seymour2000talking} emphasized alternate reasons for leaving such as faculty obsession with ``weeding out'' as opposed to supporting students and a lack of peer support. It could be that these alternative reasons (or other reasons like interest in engineering) are what motivates these students to leave the physics bachelor's degree program for engineering and further research is needed to fully understand why.

%% file: limitations.tex
A limitation of this study is that the data was confined to MSU and thus, cannot comment on how broad the findings might be regarding physics programs in general. Additionally, because the physics program at MSU is populated largely by white/Asian students (84.2\%), there isn't enough statistical power to make strong predictions about the effects of race/ethnicity in our model. We aim to to develop models with other institutional partners in order to provide discussion about the robustness of our claims. However, we find it promising that our work supports the choices made in \citet{rodriguez2016gender} given the broad differences in student populations between the two institutions.

In addition, prior preparation data (e.g., High School GPA, math placement score, etc.) was not present for all students in the study. With the aim of studying intra-STEM switching, a review of the STEM retention literature suggested that prior preparation is a key factor in STEM persistence \cite{seymour2000talking, chen2013stem}. On the other hand, some work suggests that High School GPA might not be very predictive of student outcomes at the university \cite{noble2002predicting}. To explore the impact of prior preparation, the models were analyzing with complete cases; some students had incomplete records prior to enrolling at MSU. Because of this, we are unable to comment on precisely how accurate our predictions might be regarding prior preparation.  Through our analysis, we believe that prior preparation (as we have measured it) has a small feature importance with regard to intra-STEM switching. But, we acknowledge that it could be that both physics and engineering graduates have higher than average High School GPAs and math placements and this is why we observe that these features have a small importances in our Prior preparation model.

Finally, our analysis is entirely predicated on data collected by the MSU registrar. As such, we can point to features that we find to be analytically predictive of students earning a degree in physics or engineering from MSU. However, we are unable to comment on the mechanism by which this happens or to provide any clear narratives beyond what has been presented. Qualitative work that includes interviews, focus groups, and case studies would be needed to unpack the underlying mechanisms and narratives that underpin the results presented here.

In addition to limitations within the data, the methods presented have limitations as well. First, ROC curves represent the full decision threshold space for the classification using the output probabilities from the classification model \cite{fawcett2004roc}. Thus the tails of the ROC curves may be less valuable. Additionally, we have not presented any curve fitting method in this paper for determining the optimum decision threshold via the ROC curve \cite{centor1991signal}. Instead we have opted to use visual inspection and area under the curve methods to determine if the classification model is producing believable predictions. With regards to the random forest model, random forest feature importances are relative measurements with regards to the decision trees within the forest \cite{breiman2001random}. They represent average values of changes in Gini importance and cannot be used to produced odds ratios like a logistic regression model. Additionally, for very large data sets random forests are computationally intensive. Even in this study (n=1422) the grid search was limited due to available computing resources.

The results presented suggest several implications that physics departments could put into practice. First, providing strong early motivations for taking the first modern physics course could help students decide if physics is a better choice for them as the introductory course sequence revisits most of high school physics. Second, if students come to the physics department with the express intent to switch to engineering, supporting these students' needs is likely different from supporting the needs of students who intend to stay in physics.  Lastly, performance does not appear to be a deciding factor for intra-STEM switching (at least with regard to physics and engineering). Thus, departments should place less emphasis on grades in introductory courses as a determining factor for which students to ``actively recruit'' into the major.

This work suggests several new lines of research that were not explored here but we intend to explore in the future. First, given that taking modern physics is such an important indicator for graduating with a physics degree, we intend on exploring which features are predictive of students taking this course. Second, we intend on investigating what other features characterize a student who switches to engineering or those who switch from another degree program. Third, we plan on researching what narratives underlying this intra-STEM switching can research using other methods discover as mechanisms for our observations here and in the aforementioned planned work. Finally, broadening the scope of this work, we intend on exploring which features are important for students who switch between other degree programs, leave STEM, or leave the university all together. Extending this work will provide a more complete view of the complex system students navigate from freshman year to graduation.

%% file: methods_discussion.tex
Using the predictive output of a model is important to adopt because these methods allow direct comparison to other models from different settings attempting to predict student degree outcomes in physics.  For example, the fitted model from MSU data can be directly applied to other institution data and the predictive output can be assessed. Second, a model fit on other data may provide other explanatory factors while still offering the same predictive power. In both cases this aids in verifying the explanatory features in the model as those that actually describe the system being sampled. This research provides an example, grounded in PER data and possibly a intuitive result (e.g. modern physics is the most important feature when predicting if a student will switch to engineering from a physics program), as to why prediction is important in quantitative research \cite{shmueli2010explain}. 

Figure \ref{fig:featimportance_auc} demonstrates the feature importance and ROC curve that was calculated for the Main Model. Most model's used in PER provide some feedback on the importance of each feature to the model's performance (e.g., calculating the odd's ratios for a logistic regression model's coefficients \cite{dabney2014comparative}). These feature importances give researchers a measurement that can be used to compare one feature in a model to another. In the case of logistic regression, the significance of these features is commonly evaluated via $p$-values and goodness-of-fit tests \cite{hosmer2013applied}. $p$-values have been demonstrated to be an incomplete measurement of the significance of a statistic \cite{sullivan2012using}. Goodness-of-fit tests frequently rely on residual analysis and do not always use sequestered hold out data \cite{shmueli2010explain, hosmer1997comparison, hosmer2013applied}. The above analysis used hold out data to analyze the predictive output of the models used in this paper. Hold out data provides an additional component of model analysis which helps researchers determine what the model means by examining the predictive output \cite{shmueli2010explain, hosmer1997comparison}. Analyzing a model's predictive ability does not give researchers any additional insight into explaining the system they are studying \cite{shmueli2010explain}. It does, however, provide a standard way of comparing model results. Additionally, analyzing the ROC curve is a method available to all classification models (e.g., logistic regression) and is not reserved only for Random Forest classifiers. Thus, analyzing the predictive output can be directly applied to other models published in PER in addition to the already tried and true methods of $p$-value analysis and goodness-of-fit tests.

Figure \ref{fig:aucacc} provides an analysis of the model AUC and accuracy for the Recursive model. The recursive feature elimination (RFE) method is similar in concept to methods of dimensionality reduction (e.g., factor analysis, principal component analysis (PCA)) already used in PER (e.g., \cite{scott2012exploratory}). However, unlike factor analysis and PCA, RFE allows researchers to completely exclude certain features as opposed to projecting existing features into a new space \cite{guyon2002gene}. However in some cases, researchers may see completely removing features from a model as inappropriate. Removing features could be inappropriate because there is a strong theoretical frameworks as for why a system would behave a certain way. Thus, if a model doesn't fit the data, then the model is rejected, not the data. Using RFE does not suggest that some features are theoretically inappropriate for a model. RFE is a way to measure the contribution of a feature to model prediction or goodness-of-fit without projecting the feature into a new space such as factor analysis or PCA. RFE is somewhat similar to methods already used to assess regression models in PER (i.e. explained variance in a model due to a feature \cite{potvin2016student}). However, in the case of a regression model, this typically uses an $R^2$ statistic that is calculated from the residuals used to fit the entire data set rather than on additional hold out data. There are $R^2$ statistics for regression modeling that use hold out data \cite{guyon2006introduction}; however, to the best of our knowledge, we are unaware of them being used in PER. RFE can show, relatively easily, the contribution that each feature has to the predictive power of the model (see Section \ref{sec:discussion}).

In addition to examining the model's predictive output via accuracy and AUC, the above analysis used the large data set to explore the stability of the result over the entire time domain (Figure \ref{fig:feat_over_time}). Analyzing the data for different time windows confirmed the result that taking modern physics was the most likely indicator for completing a degree in physics. The use of a direct model comparison assessing each time window's model accuracy and feature importances, provides researchers with further evidence that supports the predictive power of the model of interest.  Finally, the time domain analysis allowed for the examination of grade inflation (see Section \ref{sec:binary}) and what effect it may have had on the model. Since the total number of 3.5 or 4.0 grades grew over time, this modern approach of a sliding time window allowed us to explore the time dependence of the independent variables in relation to the model's predictive power. While not every data set in PER is as extensive as the data set used in this paper, this research demonstrates a modern way to compare complex models.

%% file: conclusions.tex
This research study has introduced to the PER community a new approach to evaluate classification models. Evaluating the predictive output of a model can provide a basis for comparing complex models across different institutional settings. In fact, the models used in the analysis can be readily made available for testing on other institutional data upon request \cite{gitrepo}.

This work attempted to take a first look at intra-STEM switching. The analysis focused on students who register as a physics major and either stay or leave the physics program for an engineering degree at Michigan State University. Using registrar data from MSU, a Random Forest classifier demonstrated that taking the first course in modern physics is a strong indicator that a student will stay in the physics program. In addition, results demonstrated that students that leave for engineering programs may ``prepare'' to do so by taking engineering courses while registered as physics majors. Finally, through this current analysis, it seems that \citet{aiken2016methods} overstated the importance of performance in introductory physics and calculus courses; grades in these courses were not of high importance in any of the models evaluated.

With the focus on assessing the quality of the predictive output of models, in this case, the ability to accurately predict who will complete a degree in physics or switch to engineering, the analysis allowed for direct comparison between contingency table results and a Random Forest model results. Summary statistics were shown to be ``significant'' via their calculated $p$-values and the size of their effect helped determine the magnitude of this significance. However, $p$-values have been demonstrated to lack the power to conclusively demonstrate significance \cite{sullivan2012using}. In this study, although taking the modern physics course was statistically significant (see Table \ref{tab:contingency_table}), the effect size was "small" and negligible for all other features. Descriptive statistics indicated that taking this course was not very important to graduating with a degree in physics. However, when used in the Random Forest classification model, it was shown that taking this first require physics course for physics majors was, intuitively, very important to predicting who will remain in the physics program or leave for an engineering degree. The model result does not invalidate the contingency table analysis. In fact, both results support each other. The contingency table demonstrates a statistical significance, while the model analysis provides its \textit{ relative} importance to other explanatory features in the data. The model also demonstrates that while the effect size is small for taking modern physics, this feature still shouldn't be dismissed from understanding why students leave for engineering. While ultimately, this problem may not have needed this sophisticated analysis to demonstrate that modern physics is important for graduating with a bachelor's degree in physics, this research has provided the basis for demonstrating the importance of model building and assessing the predictive output of complex models.